\documentclass[12pt]{article}
\usepackage{amssymb,amsfonts}
\usepackage{graphics,psboxit,amsmath}
\usepackage{subfigure}
\usepackage{graphicx}
\usepackage{verbatim}
\usepackage{color}
\usepackage{hyperref}
\hypersetup{colorlinks}

\definecolor{darkred}{rgb}{1,0,0}
\definecolor{darkgreen}{rgb}{0,0.5,0}
\definecolor{darkblue}{rgb}{0,0,1}
\definecolor{orange}{rgb}{1,0.5,0}
\definecolor{green}{rgb}{0,1,0}
\definecolor{purple}{rgb}{.5,0,1}

\hypersetup{ colorlinks,
linkcolor=darkred,
filecolor=darkgreen,
urlcolor=darkblue,
citecolor=darkblue,
linktocpage=true }

\definecolor{markcolor}{rgb}{.25,0,1}

\definecolor{markcolor2}{rgb}{1,0,0}

\definecolor{markcolor3}{rgb}{0,1,0}

\def\hybrid{\topmargin -0pt    \oddsidemargin 0.05in 
        \headheight 0pt \headsep 0pt
        \textwidth 16.5cm      
        \textheight 22,0cm       
        \marginparwidth .9in
        \parskip 5pt plus 1pt   \jot = 1.5ex}

\hybrid

\catcode`\@=11

\def\marginnote#1{}
\newcount\hour
\newcount\minute
\newtoks\amorpm
\hour=\time\divide\hour by60 \minute=\time{\multiply\hour by60
\global\advance\minute by-\hour}
\edef\standardtime{{\ifnum\hour<12 \global\amorpm={am}%
        \else\global\amorpm={pm}\advance\hour by-12 \fi
        \ifnum\hour=0 \hour=12 \fi
        \number\hour:\ifnum\minute<10 0\fi\number\minute\the\amorpm}}
\edef\militarytime{\number\hour:\ifnum\minute<10 0\fi\number\minute}

\def\draftlabel#1{{\@bsphack\if@filesw {\let\thepage\relax
   \xdef\@gtempa{\write\@auxout{\string
      \newlabel{#1}{{\@currentlabel}{\thepage}}}}}\@gtempa
   \if@nobreak \ifvmode\nobreak\fi\fi\fi\@esphack}
        \gdef\@eqnlabel{#1}}
\def\@eqnlabel{}
\def\@vacuum{}
\def\draftmarginnote#1{\marginpar{\raggedright\scriptsize\tt#1}}

\def\draft{\oddsidemargin -.5truein
        \def\@oddfoot{\sl preliminary draft \hfil
        \rm\thepage\hfil\sl\today\quad\militarytime}
        \let\@evenfoot\@oddfoot \overfullrule 3pt
        \let\label=\draftlabel
        \let\marginnote=\draftmarginnote
   \def\@eqnnum{(\theequation)\rlap{\kern\marginparsep\tt\@eqnlabel}%
\global\let\@eqnlabel\@vacuum}  }

\def\draft2{
        \def\@oddfoot{\sl preliminary draft \hfil
        \rm\thepage\hfil\sl\today\quad\militarytime}
        \let\@evenfoot\@oddfoot \overfullrule 3pt
        \let\label=\draftlabel
        \let\marginnote=\draftmarginnote
   \def\@eqnnum{(\theequation)\rlap{\kern\marginparsep\tt\@eqnlabel}%
\global\let\@eqnlabel\@vacuum}  }

\def\preprint{\twocolumn\sloppy\flushbottom\parindent 2em
        \leftmargini 2em\leftmarginv .5em\leftmarginvi .5em
        \oddsidemargin -.5in    \evensidemargin -.5in
        \columnsep .4in \footheight 0pt
        \textwidth 10.in        \topmargin  -.4in
        \headheight 12pt \topskip .4in
        \textheight 6.9in \footskip 0pt
        \def\@oddhead{\thepage\hfil\addtocounter{page}{1}\thepage}
        \let\@evenhead\@oddhead \def\@oddfoot{} \def\@evenfoot{} }

\def\numberbysection{\@addtoreset{equation}{section}
        \def\theequation{\thesection.\arabic{equation}}}

\def\underline#1{\relax\ifmmode\@@underline#1\else
        $\@@underline{\hbox{#1}}$\relax\fi}

\def\titlepage{\@restonecolfalse\if@twocolumn\@restonecoltrue\onecolumn
     \else \newpage \fi \thispagestyle{empty}\c@page\z@
        \def\thefootnote{\fnsymbol{footnote}} }

\def\endtitlepage{\if@restonecol\twocolumn \else \newpage \fi
        \def\thefootnote{\arabic{footnote}}
        \setcounter{footnote}{0}}  

\catcode`@=12 \relax

\def\figcap{\section*{Figure Captions\markboth
        {FIGURECAPTIONS}{FIGURECAPTIONS}}\list
        {Figure \arabic{enumi}:\hfill}{\settowidth\labelwidth{Figure
999:}
        \leftmargin\labelwidth
        \advance\leftmargin\labelsep\usecounter{enumi}}}
 \relax
\def\tablecap{\section*{Table Captions\markboth
        {TABLECAPTIONS}{TABLECAPTIONS}}\list
        {Table \arabic{enumi}:\hfill}{\settowidth\labelwidth{Table
999:}
        \leftmargin\labelwidth
        \advance\leftmargin\labelsep\usecounter{enumi}}}
 \relax
\def\reflist{\section*{References\markboth
        {REFLIST}{REFLIST}}\list
        {[\arabic{enumi}]\hfill}{\settowidth\labelwidth{[999]}
        \leftmargin\labelwidth
        \advance\leftmargin\labelsep\usecounter{enumi}}}
 \relax
\makeatletter
\newcounter{pubctr}
\def\publist{\@ifnextchar[{\@publist}{\@@publist}}
\def\@publist[#1]{\list
        {[\arabic{pubctr}]\hfill}{\settowidth\labelwidth{[999]}
        \leftmargin\labelwidth
        \advance\leftmargin\labelsep
        \@nmbrlisttrue\def\@listctr{pubctr}
        \setcounter{pubctr}{#1}\addtocounter{pubctr}{-1}}}
\def\@@publist{\list
        {[\arabic{pubctr}]\hfill}{\settowidth\labelwidth{[999]}
        \leftmargin\labelwidth
        \advance\leftmargin\labelsep
        \@nmbrlisttrue\def\@listctr{pubctr}}}
 \relax
\makeatother

\def\be{\begin{equation}}
\def\ee{\end{equation}}
\def\ba{\begin{eqnarray}}
\def\ea{\end{eqnarray}}

\def\a{\alpha}

\def\b{\beta}

\def\l{\lambda}

\def\s{\sigma}

\def\cN{{\cal N}}

\def\no{\noindent}

\def\IR{\relax{\rm I\kern-.18em R}}

\def\bse{\begin{small}\begin{equation*}}
\def\ese{\end{equation*}\end{small}}

\begin{document}


\renewcommand{\theequation}{\thesection.\arabic{equation}}
\csname @addtoreset\endcsname{equation}{section}

\newcommand{\eqn}[1]{(\ref{#1})}

\begin{titlepage}
\begin{center}
\strut\hfill
\vskip 1.3cm


\vskip .5in

{\Large \bf Transmission matrices in $\mathfrak{gl}_{\cN}$ \& $\mathfrak{U}_q(\mathfrak{gl}_{\cN})$ quantum\\ spin chains}

\vskip 0.5in

{\large \bf Anastasia Doikou} \vskip 0.2in

 {\footnotesize Department of Engineering Sciences, University of Patras,\\
GR-26500 Patras, Greece}
\\[2mm]
\noindent

\vskip .1cm


{\footnotesize {\tt E-mail: adoikou@upatras.gr}}\\

\end{center}

\vskip 1.0in

\centerline{\bf Abstract}

The $\mathfrak{gl}_{\cN}$ and $\mathfrak{U}_q(\mathfrak{gl}_{\cN})$ quantum spin chains in the presence of integrable
spin impurities are considered. Within the Bethe ansatz formulation, we derive the
associated transmission amplitudes, and the corresponding transmission matrices --representations of the underlying quadratic algebra--
that physically describe the interaction between the various particle-like excitations displayed by these models and the spin impurity.

\no

\vfill

\end{titlepage}
\vfill \eject


\tableofcontents

\section{Introduction}

There is a considerable amount of work devoted to the problem of integrable defects
in both quantum \cite{delmusi}-\cite{annecydef2}, and classical \cite{cozanls}-\cite{doikou-karaiskos-LL} theories.
Recently, the derivation of transmission amplitudes for the XXX and XXZ spin chains was accomplished via the
algebraic Bethe ansatz formulation \cite{doikou-karaiskos}. This was the first time that a direct computation via the Bethe ansatz equations
(BAE) was achieved in the particular framework. In this article, we focus on the
$\mathfrak{gl}_{\cN}$ and $\mathfrak{U}_q(\mathfrak{gl}_{\cN})$ quantum spin chains in the presence
of a single integrable defect and extract the physical information concerning the related
scattering processes, directly from the Bethe ansatz equations.

The algebraic formulation employed here is based on the existence of a defect Lax operator that satisfies
the same quadratic quantum algebra as the bulk monodromy matrix. Consider an one dimensional
$(N+1)$-site theory with a point-like defect on the
$n^{th}$ site; the modified monodromy matrix of the theory reads then as
\be
T(\lambda) = R_{0N+1}(\lambda)\ R_{0N}(\lambda) \ldots  L_{0n}(\lambda-\Theta) \ldots R_{01}(\lambda)\, ,
\label{basic0}
\ee
where $R$ corresponds to the ``bulk'' theory, $L$ corresponds to the defect,
and $\Theta$ is an arbitrary constant corresponding to the ``rapidity'' of the defect.
The Lax operator satisfies quadratic algebra \cite{YBE, YBE2, YBE3}
\be
R_{12}(\lambda_1 -\lambda_2)\ L_1(\lambda_1)\ L_2(\lambda_2) = L_2(\lambda_2)\ L_1(\lambda_1)\ R_{12}(\lambda_1 -\lambda_2)\, ,
\label{basicRLL}
\ee
and the $R$-matrix is a solution of the Yang-Baxter equation (see e.g. \cite{YBE, YBE2, YBE3} and references therein),
\be
R_{12}(\l_1 - \l_2)\ R_{13}(\l_1)\ R_{23}(\l_2) = R_{23}(\l_2)\ R_{13}(\l_2)\ R_{12}(\l_1- \l_2).
\ee
The monodromy matrix of the theory $T(\l)$, naturally satisfies (\ref{basicRLL}), guaranteeing the
integrability of the model. The Hamiltonian of any generic system with a point-like defect has the generic form:
\be
{\cal H} \propto - \Big ( \sum_{j\neq n=1}^{N+1}  \dot{\check R}_{j j+1}(0) + \dot{L}_{n+1 n}(0)\
L^{-1}_{n+1 n}(0) + L_{n+1 n}(0)\ \dot{\check R}_{n-1 n+1}(0)\ L^{-1}_{n+1 n}(0) \Big ), \label{hamd}
\ee
where the ``dot'' denotes the derivative with respect to the spectral parameter, also define $\check R = {\cal P}\ R$,  ${\cal P}$
is the permutation operator. Recall that the $R$ matrix reduces to the permutation operator at $\lambda =0$.

It is necessary for our purposes here to introduce some useful notation. We define the ``conjugate''
$R$-matrix via the ``crossing'' property:
\be
\bar R_{12}(\lambda) = V_1\ R_{12}^{t_2}(- \lambda - {i\cN \over 2})\ V_1
\ee
also define
\ba
&&V = \sum_{k=1}^{\cN}e_{\bar k k}, ~~~~~~~~~~~~\mathfrak{gl}_{\cN} ~~\mbox{case} \nonumber\\
&&V = \sum _{k=1}^{\cN} q^{{\cN+1 \over 2} - k} e_{\bar k k}, ~~~~\mathfrak{U}_q(\mathfrak{gl}_{\cN})
~~\mbox{case (homogenous gradation)}
\ea
where we define the ``conjugate'' index as:
\be
\bar k = \cN +1 -k, ~~~~ \mbox{and}
\ee
and the ${\cal N} \otimes {\cal N}$ matrices $e_{ij}$ with entries:
\be
(e_{ij})_{kl} = \delta_{ik}\ \delta_{jl}.
\ee

As a consequence of the Yang-Baxter equation the $\bar R$-matrix evidently satisfies:
\be
R_{12}(\lambda_1 - \lambda_2)\ \bar R_{13}(\lambda_1)\ \bar R_{23}(\lambda_2) = \bar R_{23}(\lambda_2)\
\bar R_{13}(\lambda_1)\ R_{12}(\lambda_1 - \lambda_2) \label{barR}
\ee
Moreover, we define the ``conjugate'' $L$-matrix
as:
\be
\bar L(\l) = V_1\ L_{1n}^{t_n}(-\lambda -{i\cN \over 2})\ V_1. \label{crossing}
\ee
$t_n$ denotes transposition over the ``quantum'' space $n$, the $\bar L$-matrix also satisfies the fundamental quadratic algebra (\ref{basicRLL}).
Note that we introduce the conjugate $\bar L$ matrix in order to suitably formulate the crossing property as will be transparent later in the text.
Now that the basic notions have been introduced we are ready to proceed with our main aim, which is the derivation
of the transmission matrices via the Bethe ansatz formulation.

\section{The $\mathfrak{gl}_{\cN}$ quantum spin chain}

The isotropic $\mathfrak{gl}_{\cN}$ quantum spin chain in the presence of a single point-like defect will be first investigated.
We introduce the defect $L$-matrix associated to a generic representation of $\mathfrak{gl}_{\cN}$,
which has the following simple form:
\ba
&& L(\lambda) = \lambda + i {\mathbb P}, ~~~~\mbox{and} \cr
&& \bar L(\lambda) = \lambda +{i\cN \over2}- i \bar {\mathbb P} \label{defectL}
\ea
${\mathbb P},\ \bar {\mathbb P}$ are $\cN \times \cN$ matrices expressed as:
\ba
&& {\mathbb P} = \sum_{k, l=1}^{\cN} e_{kl} \otimes P_{kl}, ~~~~P_{kl} \in \mathfrak{gl}_{\cN}, \cr
&& \bar {\mathbb P} = \sum_{k, l=1}^{\cN} e_{\bar k \bar l} \otimes P_{lk.}
\ea
The $\mathfrak{gl}_{\cN}$ elements satisfy the familiar exchange relations
\be
\Big [P_{ij},\ P_{kl} \Big ] = \delta_{il}\ P_{kj} - \delta_{kj}\ P_{il}.
\ee
The $\mathfrak{sl}_{\cN}$  algebra is generated by the elements $J^{\pm(k)},\ s^{(k)}$ defined as:
\be
J^{+(k)} = P_{k+1 k}, ~~~~J^{-(k)} = P_{k k+1}, ~~~~s^{(k)} = P_{kk} - P_{k+1 k+1}.
\ee
The latter generators satisfy:
\be
\Big [J^{+(k)},\  J^{-(k)}\Big ] = \delta_{kl}\ s^{(k)}, ~~~~~
\Big [s^{(k)},\ J^{\pm(l)} \Big ] =\pm (2 \delta_{kl} - \delta_{k l+1} - \delta_{kl-1})J^{\pm(l)}.
\ee

A generic finite-irreducible representation of the $\mathfrak{gl}_{\cN}$ algebra is associated to $\cN$ integers
$\Big (\a_1, \a_2, \ldots \a_{\cN} \Big )$ $\a_1 \geq \a_2 \geq \ldots \geq \a_{\cN}$.
Here we deal with representations that possess highest weight states such that:
\ba
&&  P_{kl}\ |\omega \rangle_n = 0 ~~~~~~k > l, ,  \nonumber\\
&&  P_{kk}\ | \omega\rangle_n = \a_k\ |\omega \rangle_n, \nonumber\\
&&  e_{kl}\ |\omega \rangle_j = 0 ~~~~~k < l, \nonumber\\
&&  e_{kk}\ |\omega \rangle_j = |\omega \rangle_j, ~~~~~~j\neq n.
\ea
The global reference state then is
\be
|\Omega \rangle = \otimes_{j=1}^{N+1} |\omega\rangle_j.
\ee
For a more detailed description see e.g. \cite{annecy1}, and references therein.

Recall also that the $R$ matrix corresponds to the fundamental representation; in this case the associated integers are
$\Big (1, 0, \ldots ,0 \Big )$. More precisely, the $R$-matrix can be written in the familiar form \cite{young}
\be
R(\lambda) = \lambda + i{\cal P}, ~~~~ {\cal P} = \sum_{k, l=1}^{\cN} e_{kl} \otimes e_{lk},
\ee
${\cal P}$ is the permutation operator associated to $\mathfrak{gl}_{\cN}$.
The spectrum and Bethe ansatz equations for this generic situation have been studied for instance in \cite{annecy1},
and the derived BAE have then the form:
\ba
&&{\mathrm X}^+_k(\lambda^{(k)} -\Theta) = -
\prod_{j=1}^{M_{k-1}} e_{-1}(\lambda^{(k)} - \lambda^{(k-1)}_j)\
\prod_{j=1}^{M_k} e_2(\lambda^{(k)} - \lambda^{(k)}_j)\
\prod_{j=1}^{M_{k+1}} e_{-1}(\lambda^{(k)} - \lambda^{(k+1)}_j) \cr
&& k \in \{1,\ 2,\ldots,\ \cN -1 \}\label{BAE1a}
\ea
here for simplicity we consider $\lambda_j^{(0)}=0$. Also, we define $M_0 = N$, $M_{\cN} =0$,
\ba
&& e_n(\lambda) = {\lambda + {in \over 2} \over \lambda -{i n \over 2}}, ~~~~~~\beta^{\pm}_k = \a_k \pm \a_{k+1}.
\cr &&
{\mathrm X}^+_k(\lambda) = e_{\beta^-_{k}}(\lambda +{i\beta_k^+ \over 2}-{i k \over 2})
\ea
it is clear that:
\be
s^{(k)}\ |\omega \rangle = \beta_k^{-}\ |\omega \rangle.
\ee
It is necessary for the purposes of this investigation --and in particular in order to confirm crossing-- to derive the BAE starting from the ``conjugate'' $L$-matrix on the defect point
(\ref{defectL}), then the BAE are given as
\be
{\mathrm X}^-_k(\lambda^{(k)}-\Theta)  = -
\prod_{j=1}^{M_{k-1}} e_{-1}(\lambda^{(k)} - \lambda^{(k-1)}_j)\
\prod_{j=1}^{M_k} e_2(\lambda^{(k)} - \lambda^{(k)}_j)\
\prod_{j=1}^{M_{k+1}} e_{-1}(\lambda^{(k)} - \lambda^{(k+1)}_j) \label{BAE2a}
\ee
where we define:
\be
{\mathrm X}^-_k(\lambda)=e_{\beta_{\cN-k}^-}(\lambda -{i\beta_{\cN-k}^+ \over 2} + {i(\cN -k) \over 2}).
\ee
Note that we have considered the generic case where $\a_1 > \a_2 > \ldots >  \a_{\cN}$. When
$\a_k= \a_{k+1}$ then the corresponding terms $e_{\beta^-_k} =1$. It will be also an instructive example to consider
the special case of the representation
characterized by integers $\a_1 = \a_2 = \ldots= \a_l = {\mathrm A} > \a_{l+1} = \ldots \a_{\cN} ={\mathrm B}$, also consider --this may be achieved using suitable shifts-- ${\mathrm A}+{\mathrm B} = l$. Then the BAE equations may be rewritten as
\ba
&& e_{{\mathrm y}}(\lambda^{(k)} -\Theta )\delta_{kl} +1 - \delta_{kl} =\cr
&& -\prod_{j=1}^{M_{k-1}} e_{-1}(\lambda^{(k)} - \lambda^{(k-1)}_j)\
\prod_{j=1}^{M_k} e_2(\lambda^{(k)} - \lambda^{(k)}_j)\
\prod_{j=1}^{M_{k+1}} e_{-1}(\lambda^{(k)} - \lambda^{(k+1)}_j) \cr
&& {\mathrm y} = {\mathrm A} - {\mathrm B} \label{BAE1b}
\ea
and the BAE associated to the ``conjugate'' $\bar L$-matrix of the special representation on the defect point (\ref{defectL}), are given as
\ba
&& e_{{\mathrm y}}(\lambda^{(k)} -\Theta ) \delta_{k \cN-l} +1 -\delta_{k \cN-l} = \cr
&& -\prod_{j=1}^{M_{k-1}} e_{-1}(\lambda^{(k)} - \lambda^{(k-1)}_j)\
\prod_{j=1}^{M_k} e_2(\lambda^{(k)} - \lambda^{(k)}_j)\
\prod_{j=1}^{M_{k+1}} e_{-1}(\lambda^{(k)} - \lambda^{(k+1)}_j) \label{BAE2b}.
\ea
Having derived the associated BAE we are now in the position to extract the relevant transmission amplitudes, and the corresponding
transmission matrices.

\subsection{The transmission matrix}
Recall that in the thermodynamic limit, which is of interest here the solutions of the BAE may be cast as strings with
a real and an imaginary part.
This is the so called string hypothesis \cite{FT, fad}, which states that the Bethe roots may be expressed as
\be
\lambda^{(n, j)} = \lambda_0 + {i \over 2} (n+1 -2j).
\ee
Also, from the asymptotic behavior of the transfer matrix one derives the quantum numbers (see also e.g. \cite{doikou-nepo1}):
\be
P_{kk} = \a_k +M_{k-1} - M_k \label{quantumn1}
\ee

The ground state as is well known consists of $\cN-1$ ``filled Fermi'' seas with real solutions, that is 1-string configurations.
The corresponding densities may be extracted from the BAE with the standard procedure \cite{FT, fad}, after taking the logarithm
and the derivative, and considering the thermodynamic limit.
Let us first introduce some notation:
\be
a_n(\lambda) = {i \over 2 \pi } {d \over d\lambda}\ln \Big (e_n(\lambda) \Big ), ~~~~
{\mathrm Y}^{\pm}_k(\l) = {i\over 2\pi} {d\over d\lambda} \ln\Big( {\mathrm X}_k^{\pm}(\lambda) \Big)
\ee
and
\be
\hat R_{jj'}(\omega) = {e^{{|\omega|\over 2}} \sinh\Big ({j_< \omega \over 2}\Big )\
\sinh \Big ((\cN - j_>){\omega \over 2}\Big )\over \sinh({\omega \over 2})\ \sinh\Big ({\cN \omega \over 2} \Big )},
~~~~~\hat a_n(\omega) = e^{-n {|\omega| \over 2}}. \label{fourier1}
\ee
see also Appendix A for more details on generic Fourier transforms.
The Fourier transforms of the ground state densities are given as:
\ba
&& \hat \sigma^{\pm(k)}_g(\omega) = \hat \sigma^{(k)}_0(\omega)  + {1\over N} \hat r_t^{\pm(k)}(\omega) \, \nonumber\\
&& \hat \sigma^{(k)}_0(\omega)= {\sinh \Big ( (\cN -k) {\omega \over 2}\Big ) \over \sinh({\cN \omega \over 2})}, ~~~~~k \in \{1,\ 2, \ldots,\ \cN -1 \}.
\label{density0}
\ea
where in the generic case described by (\ref{BAE1a}), (\ref{BAE2a}) we define (see also Appendix for generic expressions):
\be
\hat r_t^{\pm(k)}(\omega)=\sum_{j=1}^{\cN-1} \hat R_{kj}(\omega)\ \hat {\mathrm Y}^{\pm}_j(\omega).
\label{r1}
\ee
It is clear that the ``plus'' corresponds to (\ref{BAE1a}) and the ``minus'' to (\ref{BAE2a}).
In the special case described by equations (\ref{BAE1b}), (\ref{BAE2b}) we define
\ba
&& \hat r_t^{+(k)}(\omega)=  \hat R_{kl}(\omega)\ \hat a_{{\mathrm y}}(\omega) \cr
&& \hat r_t^{-(k)}(\omega)= \hat R_{k \cN-l}(\omega)\ \hat a_{{\mathrm y}}(\omega).
\label{specials}
\ea

Note that passing to the thermodynamic limit we have exploited the following basic formula in the presence of ${\mathrm m}^{(k)}$ holes in the $k^{th}$
Fermi sea:
\be
{1\over N} \sum_{j=1}^{M^{(k)}} f(\lambda_j^{(k)}) \to \int_{-\infty}^{\infty}d\lambda\ f(\lambda)\ \s^{(k)}(\lambda) -
{1\over N} \sum_{i=1}^{{\mathrm m}^{(k)}} f(\tilde \lambda_i^{(k)}).
\label{basic}
\ee
The last term corresponds to the case where ${\mathrm m}^{(k)}$ holes exist; for the ground state apparently ${\mathrm m}^{(k)}=0$.

It is then straightforward to compute via (\ref{quantumn1}), (\ref{density0}), (\ref{specials}) the ``shifted'' quantum numbers in the special cases
described by the density (\ref{specials})
(see also \cite{doikou-karaiskos} for a similar situation when computing the spin $S_z$ from the ground state):
\ba
&& P_{kk} = \tilde \a_k \cr
&& \tilde \a_k = {\mathrm A} + {l \over \cN} - 1, ~~~~~~k\leq l,  \cr
&& \tilde \a_{k} = {\mathrm B} + {l\over \cN} ~~~~~k >l \cr
&& \tilde \beta^-_k = \beta^-_k =0 ~~~~ k \neq l,~~~~~~\tilde \beta^-_l = \beta^-_l -1 = {\mathrm y} -1. \label{quantum}
\ea
Similarly, the quantum numbers associated to the defect $\bar L$ (\ref{defectL}) are given as
\ba
&& P_{kk}= \tilde \a'_{k} \cr
&& \tilde \a_k'= {\cN \over 2 }- \tilde \a_{k} =  {\cN \over 2 }- {\mathrm B} - {l \over \cN} ~~~~~k\leq \cN-l \cr
&&  \tilde \a_{k}' =  {\cN \over 2 }- \tilde \a_{k} =  {\cN \over 2 }- {\mathrm A} - {l \over \cN} +1, ~~~~k >\cN - l \
\cr
&& \tilde \beta^-_k = \beta^-_k =0 ~~~~k\neq \cN-l,~~~~~~\tilde \beta^-_{\cN-l} = \beta^-_{\cN-l} -1 = {\mathrm y} -1
\ea

Now consider the state with two holes in the first Fermi sea.
The hole in the first sea corresponds to an excitation that
carries the fundamental representation of $\mathfrak{gl}_{\cN}$ (soliton), note also that the hole in the $(\cN-1)^{th}$ sea corresponds
to an excitation that carries the anti-fundamental (conjugate) representation.
The two-hole configuration in the first Fermi sea enables the computation of the ``soliton-soliton'' $S$-matrix as
well as the computation of the associated transmission matrix.
From the BAE one derives the corresponding density of the state, which reads as
\be
\sigma^{\pm(k)}(\lambda) = \sigma_0^{(k)}(\lambda) + {1\over N} \Big ( \sum_{j=1}^2 r^{(k)}(\lambda -\tilde \lambda^{(1)}_j) +
r_t^{\pm(k)}(\lambda -\Theta) \Big ), \label{densk}
\ee
where we have used (\ref{basic}) to pass to the thermodynamic limit. We also derive the Fourier transforms:
\ba
\hat r^{(k)}(\omega) = \hat R_{k1}(\omega)\ \hat a_2(\omega) - \hat R_{k2}(\omega)\ \hat a_1(\omega), \label{densr}
\ea
with $\s_0^{(k)}(\l)$ being the densities given explicitly in (\ref{density0}),
and the Fourier transforms of $r_t^{\pm(k)}$ are defined in (\ref{r1}) and (\ref{specials}).

We are primarily interested in the densities of the first Fermi sea, which provide the scattering amplitude of the soliton-soliton
$S$-matrix and the associated transmission matrix. Recall also that
\be
\sigma^{(k)}_0(\lambda) = \varepsilon^{(k)}(\lambda), ~~~~\mbox{and}~~~~~~
\varepsilon^{(k)}(\lambda) = {1\over 2 \pi}{d p^{(k)}(\lambda)\over d \lambda} \, ,\label{EP}
\ee
with $\varepsilon^{(k)}$ and $p^{(k)}$ being the energy and the momentum of the hole
excitation in the $k^{th}$ sea, respectively. Recall also that
\be
\sigma^{(k)}(\lambda) = {1\over N}{ d h^{(k)}(\lambda) \over d\lambda} \label{sigma}
\ee
$h^{(k)}(\lambda)$ is the so-called counting function and $h^{(k)}(\tilde \lambda^{(k)}_i) = J^{(k)}_i$, where $J^{(k)}_i$ are integer numbers.

In order to identify the scattering amplitude between two holes as well as the hole-defect transmission amplitude,
we compare the expression providing the density of the first Fermi sea (\ref{densk}) with the so called quantization condition \cite{FT, Andrei-Destri}, with respect to the
hole with rapidity $\tilde \lambda_1$:
\be
\Big (e^{i N p^{(1)}(\tilde \lambda^{(1)}_1)}\ {\mathfrak S}(\tilde \lambda^{(1)}_1, \tilde \lambda^{(1)}_2, \Theta) -1 \Big )
|\tilde \lambda_1^{(1)},\
\tilde \lambda_2^{(1)}\rangle =0 \, ,
\label{QC1}
\ee
where ${\mathfrak S} = e^{i\Phi}$, and $p^{(1)}(\tilde \lambda^{(1)}_1)$ is the momentum of the respective hole in the first Fermi sea.
Comparison of the quantization condition with the state's density (\ref{densk}) would
provide the ``soliton-soliton'' scattering amplitude together with the transmission amplitude,
given that the factorization of the scattering is evident (see also \cite{doikou-nepo-mezi1, doikou-nepo-mezi2}).
The study of the one-hole state would simply provide the
transmission amplitude, which physically describes the interaction between the
particle-like excitation and the defect, thus factorization of the type: ${\mathfrak S}(\tilde \lambda_1,
\tilde \lambda_2, \Theta) = S(\tilde \lambda_1, \tilde \lambda_2)\ T(\tilde \lambda_1, \Theta)$,
in the case of the two-hole state is manifest. Keeping these considerations in mind,
the hole-hole amplitude as well as the transmission amplitude for the model with
a single defect can be derived then as
\be
S(\lambda) = \exp \Big [ - \int_{-\infty}^{\infty}\  {d\omega \over \omega} e^{-i\omega \lambda} \hat r^{(1)}(\omega)\Big ], ~~~~~
T^{\pm}(\hat \lambda)=  \exp \Big [ - \int_{-\infty}^{\infty}\  {d\omega \over \omega} e^{-i\omega \hat \lambda} \hat r_{t}^{\pm(1)}(\omega)\Big ] \label{ST}
\ee
where we have set $\lambda = \tilde \lambda_1^{(1)} - \tilde \lambda_2^{(1)}$ and $\hat \lambda = \tilde \lambda_1^{(1)} - \Theta$.

Bearing in mind the useful identity
\be
\int_0^{\infty}\ {d x \over x} e^{-{\mathrm m}x }{(1-e^{-\beta x})(1-e^{-\gamma x}) \over 1- e^{-x}} =
\ln {\Gamma({\mathrm m})\ \Gamma({\mathrm m} + \beta + \gamma) \over \Gamma({\mathrm m}+\beta)\ \Gamma({\mathrm m} + \gamma) }\, ,
\label{ident}
\ee
we may identify the scattering amplitude using (\ref{ST}):
\be
S(\lambda) = {\Gamma({i\l\over \cN} +1 )\  \Gamma(-{i\l\over \cN} +1 -{1\over \cN})\over
\Gamma(-{i\l\over \cN} +1 )\ \Gamma({i\l\over \cN} +1 -{1\over \cN})}.
\ee
Suitable string configurations may be employed in order to derive other eigenvalues of the $S$-matrix,
but we shall not give the details on this matter here, however the interested reader is referred for instance to \cite{doikou-nepo1}.
The $S$-matrix, solution of the Yang-Baxter equation may be cast as
\be
{\mathbb S}(\lambda) = {S(\lambda) \over i\l +1} \Big (i\l + {\cal P} \Big) \label{Smatrix}
\ee
recall ${\cal P}$ is the permutation operator. Notice that the eigenvalue computed from the
hole-hole interaction corresponds to the first entry of the matrix above.

The presence of one hole in the $(\cN - 1)^{th}$ sea enables the computation of the $S$-matrix
associated to the anti-fundamental representation. We avoid the details
of this computation here, but the final expression is of the form (see also \cite{doikou-nepo1})
\be
\bar S(\l) = {\Gamma(-{i\l\over \cN} + {1 \over 2})\  \Gamma({i\l\over \cN} +{1\over 2} -{1\over \cN})\over
\Gamma({i\l\over \cN} +{1\over 2} )\ \Gamma(-{i\l\over \cN} +{1\over 2} -{1\over \cN})}
\ee
and the corresponding $S$-matrix is
\be
\bar {\mathbb S}(\lambda) = {\bar S(\lambda)\over i\l +{\cN \over 2}}\Big (i\l +{\cN \over 2} - \bar {\cal P} \Big ),
~~~~~\bar {\cal P}_{12} = V_1\ {\cal P}^{t_2}\ V_1.
\ee
$\bar {\mathbb S}$ physically describes the interaction between the fundamental (a soliton, hole in the first sea) and the anti-fundamental
representation (conjugate, i.e. a hole in the $\cN-1$ sea). It is clear that the the ``conjugate'' matrix $\bar {\mathbb S}$ may be also obtained via the ${\mathbb S}$ matrix (\ref{Smatrix})
through the crossing property
\be
\bar {\mathbb S}_{12}(\l) = V_1\ {\mathbb S}^{t_2}_{12}(-\l +{i \cN \over 2})\ V_1.
\ee

The transmission amplitude in the special case (\ref{specials}) is identified via (\ref{ST}) using (\ref{ident}), and has the
following explicit form:
\be
T^+(\lambda) = {\Gamma(-{i\l \over \cN} + { \tilde \b^-_l\over2 \cN} +{l\over 2\cN})\
\Gamma({i\l \over \cN} + {\tilde \b^-_l\over 2\cN} +{\cN -l\over \cN}+{l\over 2\cN} )\over
\Gamma({i\l \over \cN} + { \tilde \b^-_l\over 2 \cN} +{l\over 2\cN})\
\Gamma(-{i\l \over \cN} + {\tilde \b^-_l\over 2\cN} +{\cN - l \over \cN}+{l\over 2\cN} )} \label{T1}
\ee
$\tilde \b_l$ are the shifted quantum numbers corresponding to the defect as derived from the ground state computation.
The generic situation can be treated in the same manner, but it is technically more intriguing due to do the various cases arising
in the analysis of the Fourier transforms, nevertheless the generalization is quite straightforward (see Appendix for more details on the related Fourier transforms).

Bearing in mind that for purely transmitting defects the following quadratic algebra is satisfied
\cite{delmusi}
\be
{\mathbb S}_{12}(\lambda_1 -\lambda_2)\ {\mathbb T}_1(\lambda_1)\ {\mathbb T}_2(\lambda_2) =
{\mathbb T}_2(\lambda_2)\ {\mathbb T}_1(\lambda_1)\ {\mathbb S}_{12}(\lambda_1 -\lambda_2)\, ,
\label{rttb}
\ee
where ${\mathbb S}$ is given in (\ref{Smatrix}), we conclude that the transmission matrix is given as
\be
{\mathbb T}(\lambda ) = {T^+(\lambda) \over i\l+ \tilde {\mathrm A} } \Big ( i\l + {\mathbb P} \Big ), ~~~~~\tilde {\mathrm A}={\mathrm A} +{l\over \cN}-1 \label{TT}
\ee
where
\be
{\mathbb P} = \sum_{k,l = 1}^{\cN-1} e_{kl} \otimes P_{kl}, ~~~~~P_{kl} \in \mathfrak{gl}_{\cN}
\ee
now the representation is characterized by the shifted quantum number $\{\tilde \a_k\}$
\footnote{Note that a legitimate shift of the generators may occur
\be
P_{kk} \to P_{kk} + c
\ee
$c$ any arbitrary constant, then the $\mathfrak{gl}_{\cN}$ algebraic relations remain intact.}.
As in the $\mathfrak{sl}_2$ case studied in \cite{doikou-karaiskos}, the transmission matrix is essentially a ``dressed'' defect matrix.

It is also useful to derive the $\bar {\mathbb T}$ matrix arising from the ``crossing property''.
This may be achieved by the second set of BAE (\ref{BAE2b}), and studying the interaction of a hole in the first sea with the defect.
The corresponding transmission amplitude arising from the interaction between the fundamental representation,
the hole in the first sea, and the defect is given by:
\be
T^-(\lambda) = \prod_{k=1}^{\cN-1}
{\Gamma(-{i\l \over \cN} +{\cN-l\over 2\cN}+ {\tilde\beta^-_{\cN-l}\over 2 \cN})\
\Gamma({i\l \over \cN} +{\cN-l\over 2\cN}+ { \tilde \b^-_{\cN-l}\over 2\cN} +{l\over \cN} )\over
\Gamma({i\l \over \cN} +{\cN-l\over 2\cN}+ {\tilde \beta^-_{\cN-l}\over2 \cN}  )\
\Gamma(-{i\l \over \cN} +{\cN-l\over 2\cN}+ { \tilde \b^-_{\cN-l}\over 2\cN} +{l\over \cN})} \label{bT1}
\ee
and the ``conjugate'' transmission matrix is then of the form:
\ba
&& \bar {\mathbb T}(\lambda) = {T^-(\lambda) \over i\lambda + {\cN \over 2}- \tilde {\mathrm B}}
\Big (i\l + {\cN \over 2} -\bar {\mathbb P}\Big ), ~~~~~\tilde {\mathrm B} = {\mathrm B} + {l \over \cN}\nonumber\\
&& \bar {\mathbb P} = \sum_{i,j} e_{\bar j\bar i} \otimes P_{ij} \label{bTT}
\ea
The latter matrix also satisfies the quadratic relation (\ref{rttb}) and is characterized by the shifted quantum numbers $\{\a_k'\}$.
Moreover it is clear that:
\be
{\mathbb T}(\lambda, \tilde {\mathrm A}, \tilde {\mathrm B}, l) = \bar {\mathbb T}(\lambda, {\cN \over 2} -\tilde {\mathrm B}, {\cN \over 2}, \tilde{\mathrm A}, \cN-l) \label{cross}
\ee
as expected from the crossing property.
Indeed, the crossing property (\ref{crossing}) effectively leads to a modification of the quantum numbers
$\a_k \to {\cN \over 2} -\a_{\cN -k+1}$. It is evident that the use of both $L,\ \bar L$ matrices is necessary in order to verify the crossing property for the derived transmission matrices as well as to further confirm the validity of our results.

\section{The $\mathfrak{U}_q(\mathfrak{gl}_{\cN})$ quantum spin chain}

The second part of the present investigation refers to the anisotropic $\mathfrak{U}_q(\mathfrak{gl}_{\cN})$ quantum spin chain
in the presence of a single defect. We introduce the $R$-matrix associated to the algebra \cite{jimbo}
\ba
&& R(\l) =\sum_{k=1}^{\cN} a(\lambda)\ e_{kk} \otimes e_{kk} + \sum_{k\neq j=1}^{\cN} b(\lambda)\ e_{kk} \otimes e_{jj} +
c \sum_{k \neq j=1}^{\cN} e^{-sign(k-j)\mu \lambda} e_{kj} \otimes e _{jk} \cr
&& a(\lambda) = \sinh(\mu(\lambda+i)), ~~~~~b(\lambda) = \sinh (\mu \lambda), ~~~~~c = \sinh(i\mu)
\ea
as well as the defect $L$-matrix and its conjugate associated to generic representations of $\mathfrak{U}_q(\mathfrak{gl}_{\cN})$,
which have the following simple forms:
\ba
L(\lambda) &=& \sum_{j=1}^{\cN} e_{jj} \otimes \sinh \Big (\mu(\lambda + i P_{ j j}) \Big )+ c
\sum_{k\neq j=1}^{\cN} e^{-sign(k-j)\mu \lambda} e_{kj} \otimes P_{kj} \cr
\bar L(\lambda) &=& \sum_{j=1}^{\cN} e_{jj} \otimes \sinh \Big (\mu(\lambda +{i\cN\over 2} -i P_{\bar j\bar j}) \Big )- c
\sum_{k\neq j=1}^{\cN} q^{k-j}e^{sign(j-k)\mu (\lambda+{i\cN \over 2})} e_{kj} \otimes P_{\bar j \bar k}
\ea
$P_{ij} \in \mathfrak{U}_q(\mathfrak{gl}_{\cN})$. We shall first introduce the $\mathfrak{U}_q(\mathfrak{sl}_{\cN})$ algebra, let
\be
a_{ij} = 2  \delta_{ij} - \delta_{i j+1}- \delta_{ij-1}
\ee
be the Cartan matrix of the $\mathfrak{sl}_{\cN}$ algebra. Also define ($q=e^{i\mu}$):
\ba
&& [m]_q = {q^m -q^{-m} \over q -q^{-1}}, ~~~~[m]_q! = \prod_{k=1}^m [k]_q, ~~~~[0]_q! =1 \nonumber\\
&&  \begin{pmatrix}
 m\cr
 n
\end{pmatrix}_q= {[m]_q \over [n]_q [m-n]_q }.
\ea
The quantum algebra $\mathfrak{U}_q(\mathfrak{sl}_{\cN})$ has the
Chevalley-Serre generators $e_i, \ f_i,\ q^{h_i \over 2}$, $i \in \{1,\ 2, \ldots, \cN\}$ that satisfy the defining relations:
\ba
&& \Big [q^{\pm h_i}, \ q^{\pm h_j} \Big ] =0, ~~~~~q^{h_i\over 2}\ e_j = a^{a_{ij}\over 2}q^{h_i\over 2}\ e_i,
~~~~q^{h_i\over 2}\ f_j = a^{-{a_{ij}\over 2}}q^{h_i\over 2}\  f_i \cr
&& \Big [ e_i,\ f_j\Big ] = \delta_{ij} {q^{h_i}- q^{-h_i}\over q-q^{-1}}
\ea
and the $q$-deformed Serre relations:
\be
\sum_{n=0}^{1 - a_{ij}} (-1)^n\begin{pmatrix}
 1- a_{ij} \cr
 n
\end{pmatrix}_q \chi_{i}^{1-a_{ij} -n}\ \chi_j\ \chi_i^n= 0,~~~~~~\chi_j \in \{e_j,\ f_j\}.
\ee
Note that $q^{h_i}=q^{P_{ii}- P_{i+1 i+1}}$, where the elements $ P_{ii} \in \mathfrak{U}_q(\mathfrak{sl}_{\cN})$.
Recall that $\mathfrak{U}_q(\mathfrak{gl}_{\cN})$ is derived by also considering the elements $q^{P_{ii}}$, so that $q^{\sum_i P_{ii}}$
belongs to the center of the algebra \cite{jimbo}. Furthermore,
there exist the elements $P_{ij} \in \mathfrak{U}_q(\mathfrak{gl}_{\cN})$, $\neq j$ such that:
\ba
&& P_{i+1 i} = e_i, ~~~~~ P_{i i+1 } = f_i \cr
&& P_{ji} = P_{ki}\ P_{jk} - q^{\mp} P_{jk}\ P_{ki}, ~~~~j \lessgtr k \lessgtr i, ~~~~~i,\ j \in \{1, 2, \ldots,  \cN \}.
\ea

As in the isotropic case a generic finite-irreducible representation is associated
to $\cN$ integers $\Big (\a_1, \a_2, \ldots, \a_{\cN} \Big )$ $\a_1 \geq \a_2 \geq \ldots \geq \a_{\cN}$
(see e.g. \cite{annecy2} and references therein).
Assume also that we deal with representations that possess highest weight states such that:
\ba
&& P_{kl}\ |\omega \rangle_n = 0, ~~~~~~k>l \nonumber\\
&& P_{kk}\ | \omega\rangle_n = \a_k\ |\omega \rangle_n \nonumber\\
&& e_{kl}\ |\omega \rangle_j = 0, ~~~~~~k<l, \nonumber\\
&& e_{kk}\ | \omega\rangle_j =  |\omega \rangle_j ~~~~~~j \neq n.
\ea
The global reference state is
\be
|\Omega\rangle = \otimes_{j=1}^{N+1} |\omega\rangle_j.
\ee

The $R$-matrix corresponds to the fundamental representation of $\mathfrak{U}_q(\mathfrak{gl})_{\cal N}$, i.e. the integers are
$\Big (1, 0, \ldots ,0 \Big )$.
The spectrum and Bethe ansatz equations for this generic situation has been studied for instance in \cite{annecy2},
and the derived BAE have then
the standard form (\ref{BAE1a}), (\ref{BAE2a}) where we now define:
\be
e_n(\lambda) = {\sinh \Big (\mu(\lambda + {in \over 2})\Big ) \over \sinh \Big (\mu(\lambda -{i n \over 2})\Big )}.
\ee
We shall also need to derive the BAE starting with the ``conjugate'' $L$-matrix on the defect point,
then the BAE are give as in (\ref{BAE2a}). Similarly, the special case where, $\a_1 = \a_2 = \ldots = \a_l ={\mathrm A} >\a_{l+1} = \ldots = \a_{\cN}= {\mathrm B}$,
with BAE (\ref{BAE1b}), (\ref{BAE2b}) is also considered.
We may now derive the relevant transmission amplitudes, and the corresponding
transmission matrices.

\subsection{The transmission matrix}

The ground state in the repulsive regime of the $\mathfrak{U}_q(\mathfrak{gl}_{\cN})$ chain as in the isotropic case,
consists of $\cN-1$ ``filled Fermi'' seas with real solutions, that is 1-string configurations \cite{doikou-nepo2}.
The corresponding density may be extracted from BAE following the standard procedure in the thermodynamic limit.
The Fourier transforms of the ground state densities are given as in (\ref{density0}), and we also define
\be
\hat R_{jj'}(\omega) = {\sinh\Big ({\nu \omega \over 2}\Big )\ \sinh\Big ({j_< \omega \over 2}\Big )\
\sinh \Big ((\cN - j_>){\omega \over 2}\Big )\over
\sinh\Big ({(\nu -1)\omega \over 2}\Big )\ \sinh({\omega \over 2})\ \sinh\Big ({\cN \omega \over 2} \Big )},
~~~~~\hat a_n(\omega) = {\sinh \Big ( (\nu - n) {\omega \over 2}\Big ) \over \sinh ({\nu \omega \over 2})}, ~~~~\nu = {\pi \over \mu}. \label{fourier2}
\ee
$0< n < 2\nu$,
(see also Appendix, and \cite{doikou-karaiskos}). Recall that passing to the thermodynamic limit we have exploited the basic formula (\ref{basic}).

It is then straightforward to compute for the special case via (\ref{quantumn1}), (\ref{density0}), (\ref{specials}) the ``shifted'' quantum numbers, starting with the $L$-matrix
(see also \cite{doikou-karaiskos}):
\ba
&& P_{kk}= {\nu \over \nu-1} \tilde \a_{k}, \cr
&& \tilde \a_k = {\mathrm A} +{l\over \cN}  -1 - {\tilde C \over \nu \cN}, ~~~~k\leq l ~~~~~\tilde C = l{\mathrm A} + (\cN - l){\mathrm B} \cr
&& \tilde \a_k = {\mathrm B} + {l \over \cN} - {\tilde C \over \nu \cN}, ~~~~k>l \cr
&& \tilde \beta^-_k =\beta^-_k=0 ~~~~~k \neq l, ~~~~\tilde \beta_l = \beta_l -1 = {\mathrm y} -1,
\ea
$\nu \over \nu -1$ is an overall renormalization factor, (see similar discussion in \cite{doikou-karaiskos}).

Similarly, the quantum numbers in the case the conjugate $\bar L$-matrix is considered, and for the special case (\ref{specials}) are given by
\ba
&& P_{kk} = {\nu \over \nu-1} \tilde \a'_k \cr
&& \tilde \a_k' = {\cN \over 2} - {\mathrm B} - {l \over \cN} +{\tilde C \over \nu \cN}, ~~~~~~k \leq \cN -l \cr
&& \tilde \a_k' = {\cN \over 2} -{\mathrm A} -{l\over \cN} +1 + {\tilde C \over \nu \cN}, ~~~~~k > \cN -l\cr
&& \tilde \beta_k = \beta_k =0 ~~~~~k\neq \cN-l, ~~~~~~\tilde \beta_{\cN-l} = \beta_{\cN-l} -1 = {\mathrm y} -1.
\ea

Now consider the state with two holes in the first Fermi sea.
The first hole as already mentioned in the isotropic case corresponds to an excitation that
carries the fundamental representation, whereas
a hole in the $(\cN-1)^{th}$ sea corresponds
to an excitation that carries the anti-fundamental (conjugate) representation, as already pointed out earlier in the text.
As in the isotropic case the two-hole configuration in the first sea enables the computation of the ``soliton-soliton'' $S$-matrix as
well as the computation of the associated transmission matrix.
From the BAE one derives the corresponding density of the state, given in (\ref{densk}), (\ref{densr}), (\ref{r1}), (\ref{specials}).
We are naturally interested in the density of the first Fermi sea. Recall also that (\ref{EP}), (\ref{sigma}) are also valid in this case.

In order to identify the scattering amplitude between two holes as well as the hole-defect transmission amplitude,
we compare the expression providing the density of the first Fermi sea (\ref{densk}) with the  quantization condition (\ref{QC1}), with respect to the
hole with rapidity $\tilde \lambda_1$. The factorization of the scattering is evident (see also \cite{doikou-nepo-mezi2})
as discussed earlier in the text.
The hole-hole amplitude as well as the transmission amplitude for the model with
a single defect can be derived then as in (\ref{ST}).

Bearing also in mind the useful identity
\be
{1\over 4}\int_{0}^{\infty} {dx\over x}\ {e^{-{\mathrm m} x}\over \sinh(x)\ \sinh(\beta x)} =
\ln \prod_{k=0}^{\infty} \Gamma({{\mathrm m}\over 2} + {\beta \over 2} + k\beta +{1\over 2}). \label{ident2}
\ee
we may identify:
\be
S(\lambda) = \prod_{k=0}^{\infty} {\Gamma(z +\cN \gamma + k \beta)\
\Gamma(z+1 +k\beta) \over \Gamma(z +\cN\gamma - \gamma + k \beta)\ \Gamma(z+\gamma+k\beta +1)}
{\Gamma(-z +\cN\gamma - \gamma + k \beta)\ \Gamma(-z+\gamma+k\beta +1) \over \Gamma(-z +\cN \gamma + k \beta)\
\Gamma(-z+1 +k\beta) }
\ee
where we define,
\be
\gamma = {1\over \nu-1},  ~~~~z = i\l \gamma, ~~~~\beta = \cN \gamma.
\ee
Suitable string configurations may be employed in order to derive other eigenvalues of the $S$ matrix,
but we shall not give the details on this matter
here, the interested reader is referred for instance in \cite{doikou-nepo2}. The $S$-matrix, solution of
the Yang-Baxter equation is then expressed as:
\ba
&& {\mathbb S}(\l) = {S(\lambda) \over {\mathrm a}(\lambda)}
\Big (\sum_{j=1}^{\cN}{\mathrm a}(\lambda)\ e_{jj} \otimes _{jj} + \sum_{k\neq j =1}^{\cN} {\mathrm b}(\lambda)\ e_{kk} \otimes e_{jj}
+ {\mathrm c} \sum_{k\neq j=1}^{\cN} e^{sign(k-j)\pi  \gamma \lambda}\ e_{kj} \otimes e_{jk} \Big )
 \cr
&& {\mathrm a}(\lambda) = \sin \Big (\pi \gamma(i\l + 1)\Big ), ~~~~~~{\mathrm b}(\lambda) = \sin (i\pi \gamma \lambda),
~~~~~{\mathrm c} = \sin(\pi \gamma). \label{Smatrix2}
\ea

Recall that the presence of one hole in the last sea enables the computation of the $S$-matrix
associated to the conjugate representation. The details
of this computation are avoided here, but the final expression is of the form (see also \cite{doikou-nepo2})
\ba
\bar S(\l) = &&\prod_{k=0}^{\infty}{\Gamma(z - \gamma + {\cN \gamma \over 2} + k \beta)\  \Gamma(z+\gamma +
{\cN \gamma \over 2} + k\beta +1)\over
\Gamma(z + {\cN \gamma \over 2} + k \beta)\  \Gamma(z +{\cN \gamma \over 2} + k\beta +1) }\cr &&
{\Gamma(-z + {\cN \gamma \over 2} + k \beta)\  \Gamma(-z +{\cN \gamma \over 2} + k\beta +1) \over
\Gamma(-z - \gamma + {\cN \gamma \over 2} + k \beta)\ \Gamma(-z+\gamma + {\cN \gamma \over 2} + k\beta +1)}
\ea
and the corresponding $S$-matrix is
\ba
&& \bar {\mathbb S}(\lambda) = {\bar S(\lambda) \over \bar {\mathrm b}(\lambda)}
\Big (\sum_{j=1}^{\cN} \bar {\mathrm a}(\lambda) e_{\bar j\bar j} \otimes e_{j j} +
\sum_{k\neq j =1}^{\cN} \bar {\mathrm b}(\lambda) e_{\bar k\bar k} \otimes e_{j j}
- {\mathrm c} \sum_{k\neq j=1}^{\cN} \tilde q^{j-k} e^{-sign(k-j) \pi \gamma (\lambda - {i\cN \over 2})} e_{\bar k\bar j} \otimes e_{kj} \Big )
\cr
&& \bar {\mathrm a}(\lambda) = \sin \Big (\pi \gamma(i\l + {\cN \over 2} - 1)\Big ),
~~~~~~\bar {\mathrm b}(\lambda) = \sin \Big (\pi \gamma (i\lambda + {\cN \over 2})\Big),~~~~~\tilde q = e^{i\pi \gamma}.
\ea
It is clear that the ``conjugate'' matrix $\bar {\mathbb S}$ may be also obtained via the ${\mathbb S}$ matrix (\ref{Smatrix2})
through the crossing property.

The transmission amplitude is identified via (\ref{ST}) using (\ref{ident2}), we are interested in the special case (\ref{specials}), and has the following explicit form:
\ba
T^+(\lambda) &=& \prod_{k=0}^{\infty}{ \Gamma(z +
{l \gamma \over 2} - {\tilde \b_l^- \gamma \over 2 } + k\beta +1)\
\Gamma(z +  {\cN\gamma \over 2} +{(\cN -l)\gamma \over 2}+ {\tilde \b_l^-\gamma \over 2 } +k\beta )  \over
\Gamma(z + {l \gamma \over 2} + {\tilde \b_l^- \gamma \over 2} +k\beta )\
\Gamma(z + {\cN \gamma\over 2} +  {(\cN-l)\gamma \over 2} - {\tilde \b_l^-\gamma \over 2} +k\beta +1)} \cr
&\times&
{\Gamma(-z + {l \gamma \over 2} + {\tilde \b_l^- \gamma \over 2} +k\beta )\
\Gamma(-z + {\cN \gamma\over 2} +  {(\cN-l)\gamma \over 2} - {\tilde \b_l^-\gamma \over 2} +k\beta +1)\over
\Gamma(-z +
{l \gamma \over 2} - {\tilde \b_l^- \gamma \over 2 } + k\beta +1)\
\Gamma(-z +  {\cN\gamma \over 2} +{(\cN -l)\gamma \over 2}+ {\tilde \b_l^-\gamma \over 2 } +k\beta )} \nonumber\\ \label{Tb}
\ea
$\tilde \b_l$ are the shifted quantum numbers corresponding to the defect as derived from the ground state computation.
Bearing in mind that the transmission matrix satisfies the quadratic algebra (\ref{rttb}),
where ${\mathbb S}$ is given in (\ref{Smatrix2}), we conclude that the transmission matrix is given as
\ba
&&{\mathbb T}(\lambda ) = {T^+(\lambda) \over \sin \Big ( \pi \gamma(i\lambda +\tilde {\mathrm A} ) \Big)} \Big (\sum_{j=1}^{\cN}
e_{jj} \otimes \sin\Big (\pi \gamma(i\lambda + P_{jj}) \Big ) +{\mathrm c}\sum_{k\neq j}^{\cN}
e^{sign(k-j)\pi \gamma \lambda} e_{kj} \otimes P_{kj} \Big ) \cr
&& \tilde {\mathrm A} = {\mathrm A} + {l\over \cN} -1 -{\tilde C \over \nu\cN}
\label{TTb}
\ea
now the representation is characterized by the shifted quantum numbers $\tilde \a_k$; all $P_{kk}$ are accordingly shifted.

It is also useful to derive the $\bar {\mathbb T}$ matrix arising from the ``crossing property''.
This may be achieved by the second set of BAE (\ref{BAE2a}) and studying the interaction of a hole in the first sea with the defect.
The corresponding transmission amplitude arising from the interaction between the fundamental representation,
the hole in the first seas and the defect is given by (in the special case (\ref{specials}):
\ba
T^-(\lambda) &=& \prod_{k=0}^{\infty}{ \Gamma(z +
{(\cN-l) \gamma \over 2} - {\tilde \b_{\cN-l}^- \gamma \over 2 } + k\beta +1)\
\Gamma(z +  {\cN\gamma \over 2} +{l\gamma \over 2}+ {\tilde \b_{\cN-l}^-\gamma \over 2 } +k\beta )  \over
\Gamma(z + {(\cN-l) \gamma \over 2} + {\tilde \b_{\cN-l}^- \gamma \over 2} +k\beta )\
\Gamma(z + {\cN \gamma\over 2} +  {l\gamma \over 2} - {\tilde \b_{\cN-l}^-\gamma \over 2} +k\beta +1)} \cr
&\times&
{\Gamma(-z + {(\cN-l) \gamma \over 2} + {\tilde \b_{\cN-l}^- \gamma \over 2} +k\beta )\
\Gamma(-z + {\cN \gamma\over 2} +  {l\gamma \over 2} - {\tilde \b_{\cN-l}^-\gamma \over 2} +k\beta +1)\over
\Gamma(-z +{(\cN-l )\gamma \over 2} - {\tilde \b_{\cN-l}^- \gamma \over 2 } + k\beta +1)\
\Gamma(-z +  {\cN\gamma \over 2} +{l\gamma \over 2}+ {\tilde \b_{\cN-l}^-\gamma \over 2 } +k\beta )} \nonumber\\  \label{bTb}
\ea
The latter matrix satisfies the quadratic relation (\ref{rttb}), and has the following explicit form:
\be
\bar {\mathbb T}(\lambda) = {T^-(\lambda) \over \sin\Big (\pi \gamma(i\lambda +{\cN \over 2}-\tilde {\mathrm B}) \Big)}
\Big (\sum_{j=1}^{\cN}
e_{j j} \otimes \sin\Big (\pi \gamma(i\lambda + {\cN \over 2}- P_{\bar j\bar j}) \Big ) -{\mathrm c} \sum_{k\neq j}^{\cN}
\tilde q^{(k-j)}e^{sign(k-j) \pi \gamma (\lambda - {i\cN \over 2})} e_{k j} \otimes P_{\bar j\bar k}\Big )
\ee
where
\be
\tilde {\mathrm B} = {\mathrm B} + {l\over \cN} - {\tilde C\over \nu \cN}. \label{bTTb}
\ee
Similarly, as in the isotropic case equation (\ref{cross}) is valid. Expressions on
Fourier transformations that provide the transmission amplitudes associated
to the generic situation may be found in the Appendix.
With this we conclude our derivation of transmission matrices in the context of
$\mathfrak{gl}_{\cN}$ and ${\mathfrak U}_q(\mathfrak{gl}_{\cN})$ quantum spin chains.

\section{Discussion}

We have been able to derive the transmission matrices for two classes of integrable spin chains associated to $\mathfrak{gl}_{\cN}$
and ${\mathfrak U}_q(\mathfrak{gl}_{\cN})$ algebras. The formulation was based on the algebraic Bethe ansatz methodology. The findings reported in the present investigation are novel in both cases, although some relevant results are known in the context of affine Toda field theories, however for a different type of defects (see e.g. \cite{corrigan-zambon2}).

We have focused here on generalizations of the so called type-II defects, which are essentially
associated to generic representations of the $\mathfrak{gl}_{\cN}$ and $\mathfrak{U}_q(\mathfrak{gl}_{\cN})$ algebras. Type-I defects, associated to generalizations of $(q)$ oscillators may be also considered in this framework. Moreover, we have restricted our investigation to representations that possess highest weight states, thus the familiar algebraic Bethe ansatz variation can be implemented. Infinite dimensional representations or representations that lack highest weight states can be also considered using a generalized Bethe ansatz  formalism together with suitable local (Darboux) gauge transformations in the spirit described in \cite{FT-XYZ, FT-XYZ2}. These and related issues will be addressed in separate investigations.

\appendix
\section{Fourier transforms}

We shall consider in this appendix the Fourier transform of the following functions in the rational case:
\be
a(x,y;\lambda) = {i\over 2 \pi}\Big ( {1\over \l + ix } - {1 \over \l + i y}\Big),
\ee
and in the trigonometric case:
\be
a(x,y;\lambda) = {i\mu \over 2 \pi} \Big ( {\cosh(\mu(\lambda +ix)) \over \sinh(\mu(\lambda +ix))}-{\cosh(\mu(\lambda +iy)) \over \sinh(\mu(\lambda +i y))}\Big ).
\ee
We distinguish three cases on the values of $x,\ y$, and we end up with the following Fourier transforms in the isotropic case:
\ba
&& \hat a(x,y;\omega)= e^{\omega x} ~~~\omega<0, ~~~~~\hat a(x,y;\omega)= e^{\omega y} ~~~~\omega>0,~~~~~x>0,~~y<0\cr
&& \hat a(x,y;\omega)= e^{\omega y} - e^{\omega x}~~~\omega > 0, ~~~~~ \hat a(x,y;\omega)= 0 ~~~~\omega < 0,~~~~~x,\ y <0 \cr
&& \hat a(x,y;\omega)= e^{\omega x} - e^{\omega y} ~~~\omega < 0, ~~~~~ \hat a(x,y;\omega)= 0 ~~~~\omega > 0,~~~~~x,\ y > 0.
\ea
In the trigonometric case, (we restrict our attention to $0 < |x|,\ |y| < \nu $, generalizations can be obtained in a straightforward manner see also \cite{doikou-karaiskos}):
\ba
&& \hat a_t(x, y; \omega) = {e^{{\nu \omega \over 2} + y \omega} - e^{-{\nu \omega \over 2} + x \omega} \over 2 \sinh({\nu\omega \over 2})},~~~~x>0,\ y<0\cr
&& \hat a(x, y; \omega) = e^{{\nu \omega \over 2}}{ e^{y \omega} - e^{x \omega} \over 2 \sinh({\nu\omega \over 2})},~~~~x,\ y < 0\cr
&& \hat a(x, y; \omega) = e^{-{\nu \omega \over 2}}{e^{x \omega} - e^{y \omega} \over 2 \sinh({\nu\omega \over 2})},~~~~x,\ y > 0.
\ea
It is clear that in  the special case $x=-y = {n\over 2}$ the expressions above reduce to the familiar Fourier transforms given in (\ref{fourier1}), (\ref{fourier2}). Moreover, in the general case we have:
\be
{\mathrm Y}_{k}^{\pm}(\lambda) = a(\a_k-{k\over 2}, \a_{k+1} -{k\over 2}; \lambda).
\ee
Having at our disposal the Fourier transforms above we obtain explicit expressions for the densities $r^{\pm(k)}$ (\ref{r1}), and hence the corresponding transmission amplitudes.

\end{document}